# Scan-less Hyperspectral Polarimetric Imaging Using Dual-Comb Single-Pixel Imaging

Shogo Tanimura[1,†], Eiji Hase[2,3,†], Hidenori Koresawa[4], Kyuki Shibuya[5], Yu Tokizane[2,3,], Akifumi Asahara[6], Kaoru Minoshima[1,6], and Takeshi Yasui[2,3,]

[1]Graduate School of Sciences and Technology for Innovation, Tokushima University, 2-1 Minami-Josanjima, Tokushima, Tokushima 770-8506, Japan

[2]Institute of Post-LED Photonics (pLED), Tokushima University, 2-1 Minami-Josanjima, Tokushima, Tokushima 770-8506, Japan

[3]Institute of Photonics and Human Health Frontier (IPHF), Tokushima University, 2-1 Minami-Josanjima, Tokushima, Tokushima 770-8506, Japan

[4]Graduate School of Advanced Technology and Science, Tokushima University, 2-1 Minami-Josanjima, Tokushima, Tokushima, 770-8506, Japan

[5]Asai Nursery, Inc., 4951 Takanoo-cho, Tsu, Mie, 514-2221, Japan

[6]Graduate School of Informatics and Engineering, The University of Electro-Communications, 1-5-1 Chofugaoka, Chofu, Japan

[†]These authors contributed equally.

## Abstract

Spectroscopic polarimetry enables detailed characterization of light-matter interactions by combining spectral and polarization information, yet conventional implementations rely on mechanical polarization modulation and are difficult to extend to fast, broadband imaging. Dual-comb spectroscopic polarimetry (DCSP) addresses these limitations by providing rapid, modulation-free access to polarization-resolved amplitude and phase spectra, but has so far been limited to point-based measurements. In this work, we introduce dual-comb spectro-polarimetric single-pixel imaging (DCSP-SPI), a new framework that combines DCSP with structured illumination and single-pixel detection to achieve scan-less hyperspectral polarimetric imaging without a camera. By exploiting the intrinsic amplitude and phase sensitivity of dual-comb interferometry, spectrally resolved images of the amplitude ratio $\psi$ and phase difference $\Delta$ are reconstructed across a broad optical spectrum. The performance of DCSP-SPI is demonstrated through imaging experiments using a birefringent resolution target and a Fabry-Pérot etalon, where frequency-dependent $\psi$ and $\Delta$ images are obtained and shown to follow the expected spectral and polarization characteristics. Furthermore, the approach is extended to hyperspectral reconstruction of full Jones matrix elements by using multiple incident polarization states. The proposed DCSP-SPI framework provides a new pathway toward scan-less, hyperspectral polarization imaging, offering a flexible platform for applications requiring quantitative spatial, spectral, and polarization analysis.

## 1. Introduction

Spectroscopic polarimetry (SP), also known as spectroscopic ellipsometry [1], is a powerful method for evaluating thin films [2], optical materials [3], and biological samples [4], since it provides access to both spectroscopic and polarization properties of light-matter interactions. Conventional SP has been realized mainly by rotating-analyzer [5], rotating-compensator [6], and phase-modulation methods [7]. While the rotating types are straightforward, they rely on mechanical polarization modulation, limiting stability, accuracy, and speed. Phase modulation with a photoelastic modulator enables faster operation, but its strong wavelength dependence hinders broadband use. In addition, all these schemes face further limitations when combined with multi-channel spectrometers, particularly in terms of measurement rate and scalability to imaging applications. Consequently, conventional SP systems struggle to simultaneously achieve high speed, broadband operation, long-term stability, and compatibility with imaging applications.

A promising approach to overcome the limitations of conventional polarization-modulation-based SP is to exploit optical frequency combs (OFCs) [8-10]. OFCs provide a unique set of characteristics, including narrow-linewidth optical modes, broad spectral coverage, equally spaced frequency components, and traceability to absolute frequency standards. Dual-comb spectroscopy (DCS) [11-14] further enhances these advantages by enabling direct acquisition of both optical amplitude and phase spectra without the need for mechanical scanning. To combine DCS with polarization-resolved measurements, dual-comb spectroscopic polarimetry (DCSP) has been proposed [15-18]. In DCSP, the amplitude and phase spectra of two orthogonal polarization components can be simultaneously obtained (namely, *x*- and *y*-polarizations in a transmission configuration, or *p*- and *s*-polarizations in an oblique-incidence reflection configuration), from which the amplitude ratio ($\psi$) and phase difference ($\Delta$) are directly derived by

$$\psi = \frac{|E_y|}{|E_x|} \quad or \quad \frac{|E_s|}{|E_p|} \tag{1}$$

$$\Delta = arg(E_y) - arg(E_x) \quad or \quad arg(E_s) - arg(E_p) \tag{2}$$

where $E_i$ denotes the complex electric-field amplitude of the corresponding polarization component. Hereafter, $\psi$ and $\Delta$ are used throughout this paper to represent the amplitude ratio and phase difference, respectively. This polarization-modulation-free scheme dramatically improves precision, speed, and stability compared with conventional methods. Accordingly, DCSP has been successfully applied thin-film characterization [15], dynamic sample measurements [16,17], nanofiber scaffold [18], and more recently, Jones-matrix retrieval [19]. If extended further to imaging, DCSP would offer even broader applicability in polarization-resolved spectroscopy and imaging science.

Although DCSP enables rapid acquisition of polarization-resolved spectra of amplitude and phase, it is inherently a point-based technique because the interferometric signals are too fast to be directly captured by conventional cameras. Thus, extending DCSP to imaging requires sequential scanning, which is time-consuming. Single-pixel imaging (SPI) [20-23] provides an attractive solution: by projecting structured light patterns and detecting the integrated response with a single photodetector, SPI reconstructs images without scanning while achieving high signal-to-noise ratio (SNR) through multiplexing strategies such as Hadamard transforms. SPI is particularly valuable in spectral regions where cameras are unavailable, such as the X-ray and terahertz regions [23-25], or in measurements of weak or ultrafast signals [23,26]. Previous studies integrated DCS with SPI (DCS-SPI) to realize scan-less hyperspectral amplitude and phase imaging [27], in which spectrally resolved optical amplitude and optical phase images of wavelength-dependent samples were successfully demonstrated. However, extending DCSP to scan-less single-pixel imaging has, to the best of our knowledge, never been reported. A key challenge is to ensure the quantitative reliability of phase-resolved imaging while simultaneously processing ultrafast polarization-resolved interferograms. Addressing this challenge motivates the present work, which aims to pioneer scan-less, spatially resolved spectro-polarimetric imaging of both $\psi$ and $\Delta$.

In this work, we propose and experimentally demonstrate a new method, dual-comb spectro-polarimetric single-pixel imaging (DCSP-SPI), which integrates DCSP with SPI. The key novelty of this approach lies in extending DCSP, originally limited to point-based measurements, to scan-less acquisition of spatially resolved spectro-polarimetric information, enabling the simultaneous retrieval of both spectroscopic and polarimetric distributions, including $\psi$ and $\Delta$. To verify the feasibility of the method, we applied DCSP-SPI to samples exhibiting both spectrally dependent responses and birefringence, and successfully obtained spatial and spectral maps of $\psi$ and $\Delta$ that showed good agreement with theoretical predictions. These results validate the proposed DCSP-SPI framework and establish its potential as a versatile platform for polarization-resolved hyperspectral imaging. We anticipate that this technique will open new possibilities for a wide range of applications, including optical component evaluation, biosensing, and advanced photonic device characterization.

## 2. Principle of operation

In this section, we outline the evolution of the experimental configuration leading to the present work, starting from a conventional DCS system and progressing through its extensions to polarization analysis and imaging. Rather than providing a full theoretical derivation, we focus on the conceptual operation and experimental configuration relevant to the proposed system. By tracing these developments, the technical distinction and novelty of the proposed DCSP-SPI approach are clarified.

**Dual-comb spectroscopy (DCS):** Figure 1(a) illustrates the basic configuration of a DCS system [14]. Two OFCs with slightly different repetition rates generate dual-comb interferograms, from which optical amplitude and phase spectra are simultaneously retrieved by Fourier transformation without mechanical scanning. From this baseline, the DCS system has diverged into two major technical extensions: polarization-resolved spectroscopy and scan-less imaging, as illustrated in Figs. 1(b) and 1(c), respectively.

**Dual-comb spectroscopic polarimetry (DCSP):** Figure 1(b) shows the configuration of DCSP, in which the detection stage of the DCS system is divided into two orthogonal polarization components, such as *x*- and *y*-polarizations (or *p*- and *s*-polarizations) [15-18]. The optical amplitude and phase spectra of each polarization are acquired simultaneously. From the ratio of the amplitude spectra, the $\psi$ spectrum is obtained, while the $\Delta$ spectrum is derived from the difference between the phase spectra of the two orthogonal polarizations, thereby enabling quantitative spectro-polarimetric analysis.

**Dual-comb spectroscopic single-pixel imaging (DCS-SPI):** A promising extension of DCS toward imaging is shown in Fig. 1(c), which depicts the DCS combined with SPI (DCS-SPI) [27]. In this configuration, the signal OFC is spatially modulated by a spatial light modulator (SLM) and projected onto the sample. The transmitted (or reflected) light is collected and detected by a single-pixel detector. By correlating the known structured illumination patterns with the corresponding detector signals, hyperspectral optical amplitude and phase images are reconstructed without spatial scanning, as demonstrated in previous studies.

**Dual-comb spectro-polarimetric single-pixel imaging (DCSP-SPI):** Figure 1(d) presents the proposed DCSP-SPI system, which integrates the principles of DCSP and DCS-SPI. The signal comb is spatially modulated by the SLM and illuminates the sample, after which the transmitted light is separated into two orthogonal polarization components. The optical amplitude and phase spectra of each polarization are simultaneously acquired, enabling the derivation of $\psi$ and $\Delta$ spectra. By correlating the structured illumination patterns with the corresponding single-pixel detector signals for both polarization channels, spatially and spectrally resolved images of optical amplitude, optical phase, and their derived polarimetric quantities are reconstructed.

This integrated approach enables scan-less spectro-polarimetric imaging, providing quantitative access to hyperspectral optical amplitude, optical phase, $\psi$, and $\Delta$ distributions, which has not been demonstrated with existing DCS-based techniques.

## 3. Methods

### 3.1 Experimental setup

A schematic of the experimental setup is shown in Fig. 2. The system consists of OFC sources, a SLM unit for single-pixel imaging, and a DCSP detection module.

**OFC sources:** A pair of mode-locked erbium-doped fiber OFCs (Neoark Co., Japan, OCLS-HSC-D100-TKSM; center wavelength = 1562 nm; spectral span = 50 nm) was used as a signal OFC and a local OFC. The repetition frequencies and carrier-envelope offset frequencies were set to $f_{rep1}$ = 100 MHz and $f_{ceo1}$ = 10.5 MHz for the signal OFC, and $f_{rep2}$ = 99.99968 MHz and $f_{ceo2}$ = 10.5 MHz for the local OFC. To realize coherent dual-comb operation, the local OFC was tightly locked to the signal OFC by finely controlling its $f_{rep2}$ with a constant repetition-rate offset of $\Delta f_{rep}$ (= $f_{rep1}$ - $f_{rep2}$ = 320 Hz). This locking was achieved using a narrow-linewidth continuous-wave (CW) laser (Redfern Integrated Optics, PLANEX; center wavelength = 1550 nm; FWHM < 2.0 kHz) as an intermediate laser [28,29]. This configuration enables coherent accumulation of interferograms in the DCSP measurement, thereby significantly enhancing the signal-to-noise ratio (SNR).

**SLM unit for single-pixel imaging:** The signal OFC was first set to vertically linear polarization using a half-wave plate (λ/2, Thorlabs, WPH05M-1550) and then directed to a reflective liquid-crystal-on-silicon spatial light modulator (LCOS-SLM, Santec, SLM100; 1050 × 1440 pixels, 10-μm pixel pitch) through a polarizing beam splitter (PBS, Thorlabs, PBS104; wavelength range: 1200–1600 nm; extinction ratio >1000) for spatial modulation. The response time of the LCOS-SLM was approximately 100 ms. The vertically polarized beam was normally incident onto the LCOS-SLM, while the liquid-crystal molecular slow axis of the SLM was oriented at 45° with respect to the incident polarization. Binary phase modulation of π/2 or 3π/2 radians was applied to the 45°-polarized component of the signal OFC by the LCOS-SLM. In combination with the PBS, this binary spatial phase modulation was converted into the corresponding binary spatial intensity modulation, which was then used to illuminate the sample. Structured illumination patterns based on Hadamard matrices were employed [27]. To reduce the effective spatial degrees of freedom and improve pattern stability, the LCOS-SLM pixels were binned to generate structured illumination patterns consisting of 32 × 32 pixels, each with an effective pixel size of 40 μm. The resulting structured illumination patterns were relay-imaged onto the sample surface through a 4f relay imaging system consisting of two lenses (L1 and L2, f = 100 mm). The polarization state of the structured illumination beam was subsequently adjusted to +5° linear polarization using an additional half-wave plate (λ/2) and a linear polarizer (P, Thorlabs, LPNIR050-MP2, extinction ratio > 100,000).

**DCSP detection:** The signal OFC transmitted through the sample was interfered with the local OFC at a beam splitter (BS, Thorlabs, BS006). To suppress aliasing in the dual-comb spectroscopy measurement, an optical bandpass filter (BPF, Thorlabs, FB-1560-12; bandwidth = 12 nm) was inserted before detection. The interfered beam was then separated into x- (horizontal) and y- (vertical) polarization components by a second polarizing beam splitter (PBS2, Thorlabs, PBS104; extinction ratio >1000). Each polarization component was detected by an individual photodetector (PD, Thorlabs, PDA10CF-EC; bandwidth = 150 MHz, wavelength range = 800–1700 nm, active area = 0.2 $mm^2$), and the corresponding interferograms were simultaneously acquired using a digitizer (National Instruments, PXI-5122). The interferogram detected in the x-polarization channel was used as the trigger signal for the digitizer. For each structured illumination pattern, 250 consecutive interferograms were coherently averaged to improve the signal-to-noise ratio (SNR). To investigate the effect of coherent accumulation on image quality, additional measurements were performed with coherent accumulation numbers of 1 and 10, and the corresponding image datasets were reconstructed. Only the center-burst region of each interferogram, consisting of 20,000 data points (6.4% of one interferogram period, 312,499 data points), was extracted. The truncated interferograms were then zero-filled to compensate for the missing time-domain segments, followed by Fourier transformation to retrieve the optical amplitude and phase spectra for each polarization component. The Fourier-transformed spectra had a physical spectral resolution of 1.562 GHz (12.5 pm at 1560 nm) and a display spectral resolution of 100 MHz, the latter being achieved by zero-padding without improving the actual spectral resolution. From the optical amplitude and phase spectra of the orthogonal polarization components, the $\psi$ and $\Delta$ spectra were independently calculated. To reconstruct a 32

× 32-pixel image, a total of 1024 structured illumination patterns were sequentially projected and the corresponding interferograms were recorded. The resulting $\psi$ and $\Delta$ spectra were then correlated with the corresponding structured illumination patterns to reconstruct spatially resolved spectro-polarimetric images.

### 3.2 Sample

Two types of samples were used in this study, either individually or in combination, to evaluate the spectro-polarimetric imaging performance of the proposed DCSP-SPI system in terms of polarization sensitivity, spatial resolution, and spectral resolution.

**Birefringent resolution target:** A birefringent resolution target was used to assess the polarization-analysis capability and spatial resolution. A 1951 USAF birefringent resolution target (Thorlabs, R3L1S1B) fabricated from a molecularly aligned liquid-crystal polymer was employed. The target exhibited a nominal retardance of 280 nm and a spatially varying orientation of the fast axis. The fast-axis orientation was set to 60° with respect to the sample horizontal axis within the patterned regions corresponding to the USAF elements, while the surrounding background region had a fast-axis orientation of 0°, i.e., parallel to the sample horizontal axis. This sample enabled quantitative evaluation of the system's ability to resolve spatial variations in $\psi$ and $\Delta$ arising from birefringence.

**Fabry-Pérot etalon filter:** We used a Fabry-Pérot etalon plate (Koshin Kogaku Co., Ltd., Japan, Etalon Filter-90GHz) with a center wavelength of 1550 nm, a free spectral range (FSR) of 90 GHz, and a finesse of 8.5 to evaluate the spectral performance of the system. This sample provided well-defined spectral modulation, allowing verification of the spectrally resolved amplitude and phase imaging capability of the DCSP-SPI system.

In some experiments, the birefringent resolution target and the etalon filter were stacked to simultaneously introduce spatially varying polarization properties and wavelength-dependent spectral features, thereby enabling comprehensive evaluation of the system's combined spectroscopic, polarimetric, and imaging performance.

### 3.3 Use of artificial intelligence

During manuscript preparation, ChatGPT (OpenAI) was used to assist with English language editing, improvement of sentence structure, and refinement of the presentation. The authors critically reviewed and revised all AI-assisted text and take full responsibility for the final content. Artificial intelligence was not used for experimental design, data analysis, interpretation of results, or scientific decision-making.

## 4. Results

### 4.1 Characterization of samples

To evaluate the polarization analysis performance of the developed system, point measurements were first conducted using the 1951 USAF birefringent resolution target. For each measurement, 250 consecutive interferograms were coherently averaged to improve the signal-to-noise ratio (SNR). The coherently averaged interferograms were processed with a physical spectral resolution of 1.562 GHz and an interpolated display spectral resolution of 100 MHz. From these data, the polarization analysis results at 192.0125 THz were extracted, as presented below.

Figure 3(a) shows the sample rotation angle dependence of $\psi$. The results obtained from the patterned birefringent region are plotted in red, while those from the background birefringent region are shown in blue. A clear periodic dependence on the sample rotation angle is observed in both regions. For comparison, their corresponding theoretical curves are also plotted in solid red and blue lines, and excellent agreement between the experimental results and theory is confirmed. Figure 3(b) presents the sample rotation angle dependence of $\Delta$. Similar to the behavior of $\psi$, a distinct rotation-angle-dependent behavior is observed for the patterned region (red plots) and the background region (blue plots). The experimentally obtained phase difference is in good agreement with the theoretical prediction (red and blue lines). The difference between the patterned and background regions in both $\psi$ and $\Delta$ appears as a horizontal shift in the rotation-angle dependence, which originates from the 60° difference in the fast-axis orientation between the two regions. The measured characteristics are well reproduced by the theoretical model using a retardance of 280 nm. These results verify the accuracy of the developed DCSP system in point-based spectro-polarimetric measurements, providing the basis for its subsequent extension to single-pixel imaging.

Next, the spectral performance of the system was evaluated using a Fabry–Pérot etalon filter. Figure 3(c) shows the measured transmission spectrum, in which periodic spectral interference fringes characteristic of the etalon are clearly observed. Figure 3(d) shows an enlarged view of the spectrum, from which a free spectral range (FSR) of 90 GHz and a finesse of 8.3 are obtained, in good agreement with the specifications of the etalon.

From these results, it is confirmed that both polarization analysis and spectroscopic measurement can be performed accurately in point measurements using the developed system.

### 4.2 Single-frequency polarimetric single-pixel imaging

Single-frequency polarimetric SPI was performed using the birefringent resolution target to demonstrate the polarimetric imaging capability of the proposed DCSP-SPI system. First, a $\psi$ image at 192.0125 THz was reconstructed with an image size of 1.28 mm × 1.28 mm and a spatial sampling of 32 × 32 pixels, corresponding to an effective pixel size of 40 μm. Figures 4(a) and 4(b) show the reconstructed $\psi$ images when the sample was rotated by 10° and 45°, respectively. In both cases, the test pattern is clearly visualized by contrast in the amplitude ratio $\psi$. A negative pattern is observed in Fig. 4(a), whereas a positive pattern appears in Fig. 4(b). This contrast reversal is consistent with the change in the relative orientation between the fast axis of the birefringent elements and the incident +5° linear polarization. As predicted by the point-measurement results in Fig. 3(a), the amplitude ratio $\psi$ changes from values below unity to above unity with increasing sample rotation angle, leading to the observed transition from negative to positive image contrast. These results confirm the validity of the reconstructed $\psi$ images. The red plots in Figs 4(c) and 4(d) show the $\psi$ profiles extracted along the red lines indicated in Figs. 4(a) and 4(b), respectively.

Clear step-like profiles corresponding to the patterned and background regions are observed. From these profiles, the top and bottom amplitude ratio values were quantitatively determined to be 0.2740 ± 0.029 and 0.1068 ± 0.034 for the sample rotated by 10°, and 0.4312 ± 0.045 and 0.1296 ± 0.038 for the 45°-rotated sample, where the values represent the mean ± standard deviation. For comparison, the predicted $\psi$ values for the patterned and background regions, calculated from the sample parameters, are overlaid as blue solid and dashed lines (0.2918 and 0.1281 for the 10°-rotated sample, and 0.1280 and 0.4185 for the 45°-rotated sample). Although the spatial resolution is limited by the finite spatial sampling of the structured illumination patterns, the measured profiles show good agreement with the predicted values.

Next, a phase difference $\Delta$ image at 192.0125 THz was reconstructed with the same image size and pixel sampling. Figures 4(e) and 4(f) show the reconstructed $\Delta$ images for sample rotated angles of 10° and 45°, respectively. In both cases, the test pattern is clearly visualized by contrast in the phase difference. The circular feature observed in the background region suggests that the measured $\Delta$ is influenced by residual spherical wavefront components in the optical system. The red plots in Figs. 4(g) and 4(h) present the $\Delta$ profiles extracted along the red lines in Figs. 4(e) and 4(f), respectively, where clear step-like profiles corresponding to the patterned and background regions are again observed. From these profiles, the top and bottom $\Delta$ values were quantitatively determined to be 1.6636 ± 0.1502 ° and 0.0575 ± 0.2755 ° for the 10°-rotated sample, and 1.7459 ± 0.3324 ° and -0.4708 ± 0.3204 ° for the 45°-

rotated sample. For comparison, the predicted $\Delta$ values for the patterned and background regions, derived from the sample parameters, are overlaid as blue solid and dashed lines (1.5481° and 0.5314° for the 10°-rotated sample, and -0.5046° and 1.2479° for the 45°-rotated sample). The experimentally measured $\Delta$ profile in Fig. 4(g) show good agreement with the predicted values. In contrast, the larger deviation observed for the background region in Fig. 4(h) is consistent with the residual spherical wavefront component identified in Fig. 4(f).

Overall, these results demonstrate that the proposed DCSP-SPI system enables scan-less polarimetric imaging with quantitative reconstruction of both $\psi$ and $\Delta$, and that the observed agreement between experiment and theory confirms the validity and robustness of the proposed imaging approach, despite the limited spatial resolution imposed by the number of spatial modulation patterns.

### 4.3 Spectro-polarimetric single-pixel imaging

To evaluate the spectro-polarimetric imaging capability of the proposed DCSP-SPI system, a birefringent resolution target was combined with a Fabry–Pérot etalon filter and used as a sample. The birefringent resolution target was fixed without rotation, such that its fast axis was oriented at an angle of 60°. From this dataset, $\psi$ and $\Delta$ images were extracted at four representative optical frequencies: $f_1$ = 191.9521 THz, $f_2$ = 191.9931 THz, $f_3$ = 192.0421 THz, and $f_4$ = 192.0831 THz. Among them, $f_1$ and $f_3$ correspond to transmission peaks of the etalon filter, whereas $f_2$ and $f_4$ correspond to transmission minima, as shown in Fig. 5(a).

Figures 5(b)-(e) show the $\psi$ images obtained at $f_1$, $f_2$, $f_3$, and $f_4$, respectively. The image quality of the $\psi$ images varies depending on the etalon transmission, indicating that the reconstructed images accurately reflect the spectral characteristics of the sample. At the transmission-peak frequencies $f_1$ and $f_3$, the $\psi$ images clearly reproduce the features of the birefringent resolution target. At the transmission-minimum frequencies $f_2$ and $f_4$, although the image SNR is reduced, the $\psi$ images still exhibit spatial distributions consistent with those observed at $f_1$ and $f_3$. Figures 5(f)-(i) present the corresponding $\Delta$ images at $f_1$, $f_2$, $f_3$, and $f_4$. Similar to the $\psi$ images, the SNR of the $\Delta$ images changes in accordance with the etalon transmission, again confirming consistency with the spectral response of the sample. The $\Delta$ images at $f_1$ and $f_3$ clearly capture the birefringent features of the resolution target, while those at $f_2$ and $f_4$ show reduced SNR but retain the same qualitative spatial characteristics as the peak-frequency images. Unlike the $\psi$ images, where reduced SNR mainly degrades the image contrast, the corresponding $\Delta$ noise appears as spatially random fluctuations over the wrapped phase range ($-\pi$ to $\pi$), reflecting the fundamentally different noise characteristics of amplitude- and phase-based imaging.

The representative images in Fig. 5 provide snapshots of the hyperspectral dataset at selected optical frequencies. The complete dataset consists of 10,000 pairs of spectrally resolved $\psi$–$\Delta$ images acquired over the full spectral range of 191.5–192.5 THz with a physical resolution of 1.562 GHz and a display resolution of 100 MHz. To visualize the continuous spectral dependence, the evolution of the $\psi$ and $\Delta$ image pairs as a function of optical frequency is shown in Movie 1, covering the spectral range from 191.9049 to 192.0421 THz. The movie demonstrates the correlated spectral evolution of $\psi$ and $\Delta$ images, reflecting both the transmission profile of the Fabry-Pérot etalon and the birefringent structure of the resolution target, thereby highlighting the hyperspectral capability of DCSP-SPI. These results demonstrate the capability of DCSP-SPI for scan-less quantitative spectro-polarimetric imaging over a continuous optical spectrum, providing simultaneous access to the spectral evolution of birefringent contrast through the hyperspectral $\psi$ and $\Delta$ images.

## 5. Discussion

The experimental results presented in Sections 4.2 and 4.3 demonstrate the feasibility of the proposed DCSP-SPI scheme for quantitative spectro-polarimetric imaging. In this section, we discuss practical considerations for improving the acquisition speed and extending the functionality of the system.

### 5.1 Image acquisition time

Although the proposed DCSP-SPI scheme enables scan-less spectro-polarimetric imaging without a camera, the image acquisition time remains a critical performance metric for practical applications. Here, we quantitatively evaluate the image acquisition time of the present system and discuss strategies for further acceleration. The image acquisition time of DCS-SPI is determined by several key factors: the accumulation number of interferograms per illumination pattern, the pattern switching time of the LCOS-SLM, the acquisition time of each dual-comb interferogram, and the total number of illumination patterns. In the present implementation, as highlighted in red in the upper part of Table 1, the measurements employed a coherent accumulation number of 250, an individual interferogram acquisition time of 9.5 ms, an accumulated interferogram acquisition time of 2.4 s, and a pattern switching time of 2.2 s, all per structured illumination pattern, with a total of 1024 structured illumination patterns. The pattern switching time is governed by the temporal response of the LCOS-SLM, whereas the interferogram acquisition time is determined by $\Delta f_{rep}$ and the associated data transfer time. The number of illumination patterns is directly related to the image pixel number, which was set to 1024 pixels in this study. Under the present experimental conditions, the total image acquisition time amounts to 4710 s. Although the imaging process is scan-less, this acquisition time is not sufficiently fast for many practical applications.

Without changing the hardware configuration of the present system, the most straightforward approach to reduce the acquisition time is to decrease the accumulation number. Figures 6(a) and 6(b) compare the reconstructed $\psi$ and $\Delta$ images obtained with coherent accumulation numbers of 1, 10, and 250. Although reducing the accumulation number increases image noise and decreases image contrast, the birefringent target remains clearly distinguishable even under single-shot acquisition. These results indicate that the coherent accumulation number can be substantially reduced without losing the essential birefringent features, thereby enabling a significant reduction in the interferogram acquisition time. Nevertheless, as highlighted in blue in the upper part of Table 1, simply reducing the number of coherent accumulations is insufficient to substantially shorten the total acquisition time because the slow pattern switching of the LCOS-SLM, rather than interferogram acquisition, dominates the overall measurement time.

In the present implementation, as long as the LCOS-SLM is employed, further acceleration is primarily limited by its pattern switching speed. To overcome this limitation, an SLM with a much faster switching speed is required. Replacing the LCOS-SLM with a faster SLM shifts the dominant bottleneck to interferogram acquisition, reducing the estimated total measurement time. A promising candidate is a digital micromirror device (DMD), which can achieve pattern switching times as short as 240 μs per pattern using commercially available devices. The estimated acquisition times for different coherent accumulation numbers using the DMD-based implementation are summarized in the lower part of Table 1. In this case, the pattern switching time becomes shorter than the interferogram acquisition time, and the total image acquisition time is primarily determined by the latter. As a result, a substantial reduction in acquisition time can be achieved by decreasing the accumulation number. For an accumulation number of 250, the reduction is modest, yielding approximately a factor-of-two improvement compared with the LCOS-SLM case (from 4710 s to 2458 s). In contrast, for accumulation numbers of 10 and 1, the dependence on the pattern switching time is effectively eliminated, enabling a dramatic reduction in the total acquisition time to the order of 100 s and 10 s, respectively.

An additional and complementary strategy is to reduce the number of projected illumination patterns by employing compressed sensing. While conventional Hadamard SPI generally requires one illumination pattern for each image pixel, compressed sensing enables accurate image reconstruction from substantially fewer measurements by exploiting signal sparsity. Because the achievable compression ratio depends strongly on the sample characteristics, reconstruction algorithm, and acceptable reconstruction error, quantitative evaluation is beyond the scope of the present work. Nevertheless, compressed sensing has been successfully applied to single-pixel complex-amplitude imaging and holography [30,31], suggesting that the same strategy can be extended to the proposed DCSP-SPI framework. By combining fast pattern modulation, reduced coherent accumulation, and compressed sensing, the proposed framework offers a practical route toward further reducing the acquisition time.

### 5.2 Toward hyperspectral Jones matrix imaging

Beyond enabling scan-less hyperspectral imaging, the proposed DCSP-SPI introduces a unified framework for quantitative polarization imaging by integrating dual-comb spectroscopic polarimetry with single-pixel imaging. Unlike conventional polarization imaging techniques, which generally rely on sequential polarization modulation or multiple image acquisitions, DCSP-SPI concurrently acquires quantitative hyperspectral amplitude-ratio ($\psi$) and phase-difference ($\Delta$) distributions from mutually coherent dual-comb interferograms without mechanical polarization modulation. Consequently, the spatial, spectral, and polarization information are intrinsically synchronized within a single measurement, providing quantitative access to wavelength-dependent anisotropy that is difficult to achieve with conventional polarization imaging techniques. This simultaneous acquisition of hyperspectral $\psi$ and $\Delta$ distributions constitutes one of the principal advantages of the proposed framework.

An important extension of this capability is hyperspectral Jones matrix imaging. While $\psi$–$\Delta$ imaging provides direct access to polarization amplitude ratio and phase difference, reconstruction of the full complex-valued Jones matrix enables complete characterization of polarization-dependent optical responses, including amplitude transmission, phase retardation, and polarization coupling. Because DCSP directly retrieves the complex optical fields of two orthogonal polarization components, integration with Jones matrix analysis naturally extends the present framework from $\psi$–$\Delta$ imaging to hyperspectral Jones matrix imaging without mechanical polarization modulation [19]. This capability further expands the measurement space from spectro-polarimetric imaging to complete hyperspectral polarization characterization.

As a proof-of-concept demonstration, hyperspectral Jones matrix imaging was implemented by combining two independently acquired DCSP-SPI datasets obtained with incident linear polarization states of 5° and 95°, respectively. Since reconstruction of the full 2 × 2 Jones matrix requires polarization analysis under at least two known incident polarization states, the measured output polarization states were combined with the corresponding incident states through inverse matrix operations [19]. A birefringent resolution target stacked with a Fabry–Pérot etalon filter was used as a sample to demonstrate the feasibility of hyperspectral Jones matrix imaging. The birefringent resolution target was fixed without rotation such that the fast axis of the background region was oriented at 45°, whereas that of the target region was oriented at 105°.

Figure 7 shows a representative set of spectrally resolved Jones matrix images at an optical frequency of 192.0421 THz, corresponding to a transmission peak of the Fabry–Pérot etalon [see $f_3$ in Fig. 5(a)]. The reconstructed dataset consists of the real and imaginary parts of the four Jones matrix elements: $J_{00}$ ($J_{00r}$ and $J_{00i}$), $J_{01}$ ($J_{01r}$ and $J_{01i}$), $J_{10}$ ($J_{10r}$ and $J_{10i}$), and $J_{11}$ ($J_{11r}$ and $J_{11i}$), shown in panels (a)–(h), respectively. The color scale of all images is normalized to $J_{00r}$. Although the present implementation serves as a proof-of-concept demonstration, the reconstructed Jones matrix images exhibit distinct spatial contrasts reflecting the amplitude and phase responses of the birefringent sample. The continuous spectral evolution of these Jones matrix images over the measured optical frequency range is provided in Movie 2, illustrating the hyperspectral imaging capability of the proposed framework.

The reconstructed Jones matrix can be validated by considering the expected response of an ideal linear retarder. The birefringent resolution target used in this study can be regarded as a spatially distributed linear retarder whose fast-axis orientation differs by 60° between the background and target regions. When the fast axis is aligned with the measurement coordinate system, the Jones matrix is diagonal. In contrast, when the fast axis is rotated with respect to the measurement coordinates, as in the present sample, both diagonal and off-diagonal elements appear. For an ideal linear retarder under this condition, the off-diagonal elements satisfy the symmetry relation $J_{01} = J_{10}$, providing a simple experimental criterion for validating the reconstructed Jones matrix.

Based on the expected response of a linear retarder, the experimentally reconstructed Jones matrix was evaluated by comparing the diagonal and off-diagonal components. In both the background and target regions, the off-diagonal elements, $J_{01}$ and $J_{10}$, exhibit nearly identical values, although a slight discrepancy is observed in the background region. This agreement supports the validity of the reconstructed Jones matrix and is consistent with the expected response of a spatially distributed linear retarder. Furthermore, the magnitude and sign of the off-diagonal components differ clearly between the background and target regions, in good agreement with the designed 60° difference in the fast-axis orientation. The off-diagonal components are also considerably smaller than the diagonal components, and only a small difference in the Jones matrix values is observed between the background and target regions, reflecting the relatively weak birefringence of the sample. This observation is consistent with the retardance of the birefringent resolution target, which is approximately $\pi/6$ at the operating wavelength.

Overall, these results indicate that the principal polarization characteristics of the birefringent sample are successfully captured by the reconstructed Jones matrix. Nevertheless, some limitations remain in the current implementation. In particular, the diagonal components exhibit less distinct target patterns than the off-diagonal components. This limitation is attributed not only to the self-triggered interferogram acquisition but also to the reconstruction procedure of the Jones matrix itself. Because the Jones matrix is reconstructed by combining two independently acquired datasets corresponding to different incident polarization states, small differences in the temporal reference between the measurements are not canceled as common-mode errors. Consequently, phase inconsistencies accumulate during the inverse matrix reconstruction, reducing the reconstruction fidelity, particularly for the diagonal components. This limitation is expected to be mitigated by introducing an external reference trigger, which would provide a common temporal reference for all polarization measurements, suppress measurement-to-measurement phase inconsistencies, and improve the reconstruction fidelity of the full Jones matrix. Furthermore, the polarization-controlled pulse train proposed in our previous DCSP framework [19] provides a promising route toward simultaneous acquisition of multiple incident polarization states within a single measurement. Such an approach would eliminate measurement-to-measurement temporal inconsistencies and enable one-shot hyperspectral Jones matrix imaging while preserving the scan-less nature of the proposed DCSP-SPI framework. More importantly, the present results demonstrate that dual-comb spectroscopic polarimetry can be naturally extended from point-based measurements to scan-less hyperspectral imaging while preserving quantitative polarization information. This establishes a unified measurement framework that provides spatial, spectral, and polarization-resolved information in a single measurement, paving the way toward practical real-time hyperspectral Jones matrix imaging when combined with the acquisition-speed improvements discussed in Section 5.1.

## 6. Conclusion

In this work, we proposed and experimentally demonstrated DCSP-SPI, which integrates dual-comb spectroscopic polarimetry with single-pixel imaging. The proposed approach enables scan-less quantitative spectro-polarimetric imaging without the use of a camera, providing spatially and spectrally resolved maps of the amplitude ratio $\psi$ and phase difference $\Delta$. The validity of DCSP-SPI was confirmed through quantitative agreement with point-based DCSP measurements and by spectro-polarimetric imaging experiments using birefringent samples combined with a Fabry–Pérot etalon. Hyperspectral $\psi$ and $\Delta$ images, together with their continuous spectral evolution, were successfully reconstructed, demonstrating that DCSP-SPI can simultaneously capture spatial and spectral polarization information in a hyperspectral imaging framework.

We further discussed the image acquisition time and identified the pattern switching speed of the LCOS-SLM as the dominant bottleneck in the present implementation. Although the present implementation was designed as a proof-of-principle demonstration, our analysis shows that substantial acceleration is feasible by employing faster SLMs such as DMDs and by reducing the number of illumination patterns through compressed sensing strategies, paving the way toward acquisition times on the order of seconds.

Finally, hyperspectral single-pixel imaging of full Jones matrix elements was demonstrated using a two-step measurement scheme with two independent incident polarization states. Although this proof-of-concept implementation demonstrates the feasibility of hyperspectral Jones matrix imaging, the use of polarization-controlled pulse sequences offers a promising route toward single-step acquisition of full Jones matrix information [19]. Such an extension would further simplify the measurement procedure and significantly enhance the temporal efficiency of scan-less, fully polarization-resolved hyperspectral imaging. Overall, the proposed DCSP-SPI framework establishes a versatile and scalable platform for quantitative polarization imaging, with broad potential applications in optical component characterization, anisotropic material analysis, biosensing, and advanced photonic device evaluation.

## Data Availability

The data that support the findings of this study are available from the corresponding author upon reasonable request.

## Funding

This work was supported in part by the Japan Society for the Promotion of Science (JPJS00420240022, 24K21237).

## Conflict of Interest

The authors have no conflicts of interest to disclose.

## Author Contributions

**Shogo Tanimura:** Investigation (equal); Data curation (lead); Methodology (supporting); Writing - review & editing (supproting). **Eiji Hase:** Investigation (equal); Methodology (lead); Writing - review & editing (equal). **Hidenori Koresawa:** Methodology (supporting). **Kyuki Shibuya:** Methodology (supporting). **Yu Tokizane:** Supervision (supporting). **Akifumi Asahara:** Methodology (supporting). **Kaoru Minoshima:** Methodology (supporting). **Takeshi Yasui:** Conceptualization (lead); Project administration (lead); Methodology (equal); Writing - original draft (lead); Writing - review & editing (equal); Supervision (lead); Funding acquisition (lead).

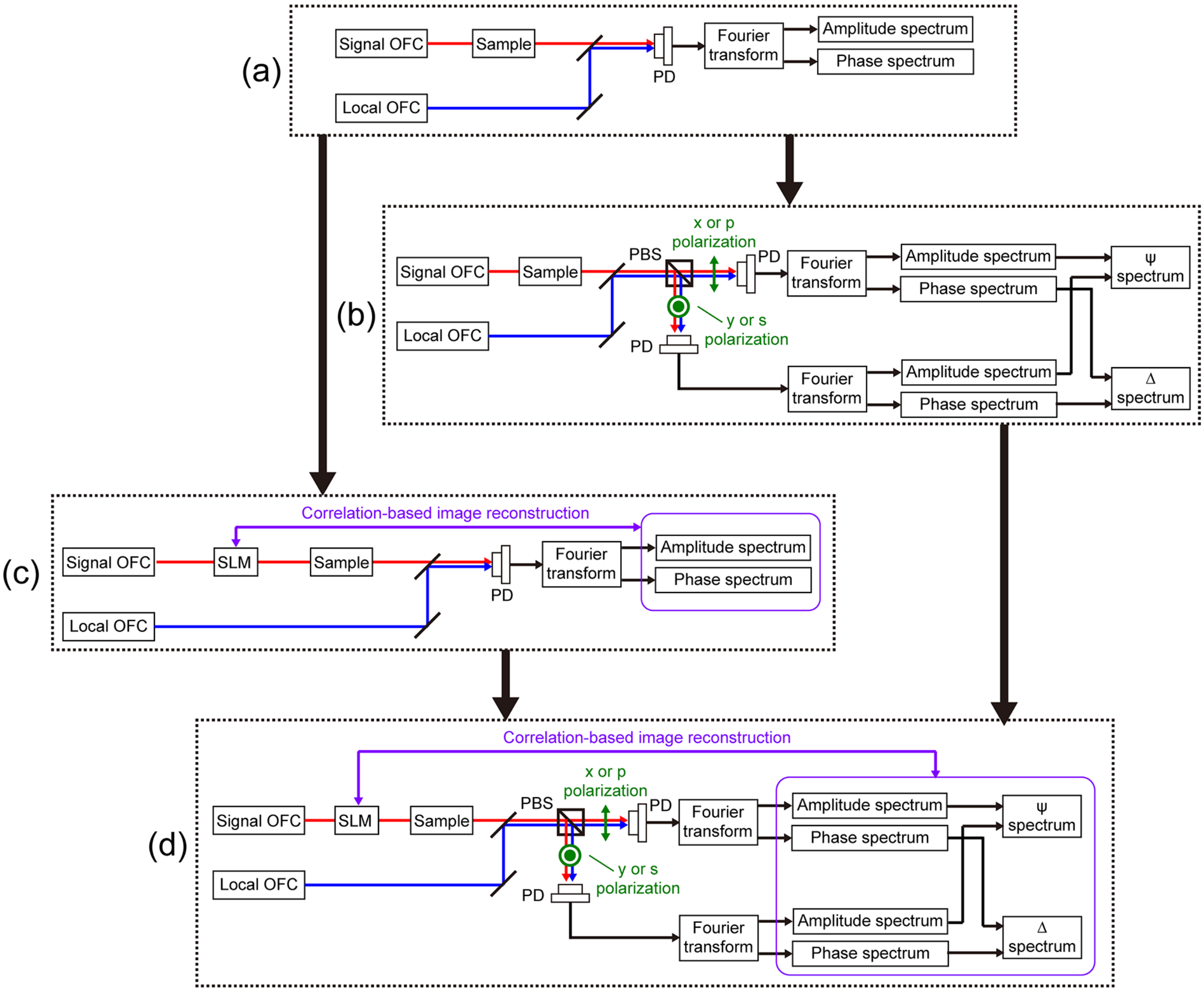

Fig. 1. Evolution of dual-comb spectroscopy toward scan-less spectro-polarimetric imaging. (a) Conventional dual-comb spectroscopy (DCS), enabling retrieval of optical amplitude and phase spectra. (b) Dual-comb spectroscopic polarimetry (DCSP), enabling measurement of polarization-resolved amplitude and phase spectra for deriving the amplitude ratio $\psi$ and phase difference $\Delta$. (c) Dual-comb spectroscopic single-pixel imaging (DCS-SPI), extending DCS to scan-less hyperspectral amplitude and phase imaging through correlation-based image reconstruction. (d) Proposed dual-comb spectro-polarimetric single-pixel imaging (DCSP-SPI), integrating DCSP with DCS-SPI to achieve scan-less hyperspectral imaging of optical amplitude, optical phase, $\psi$, and $\Delta$.

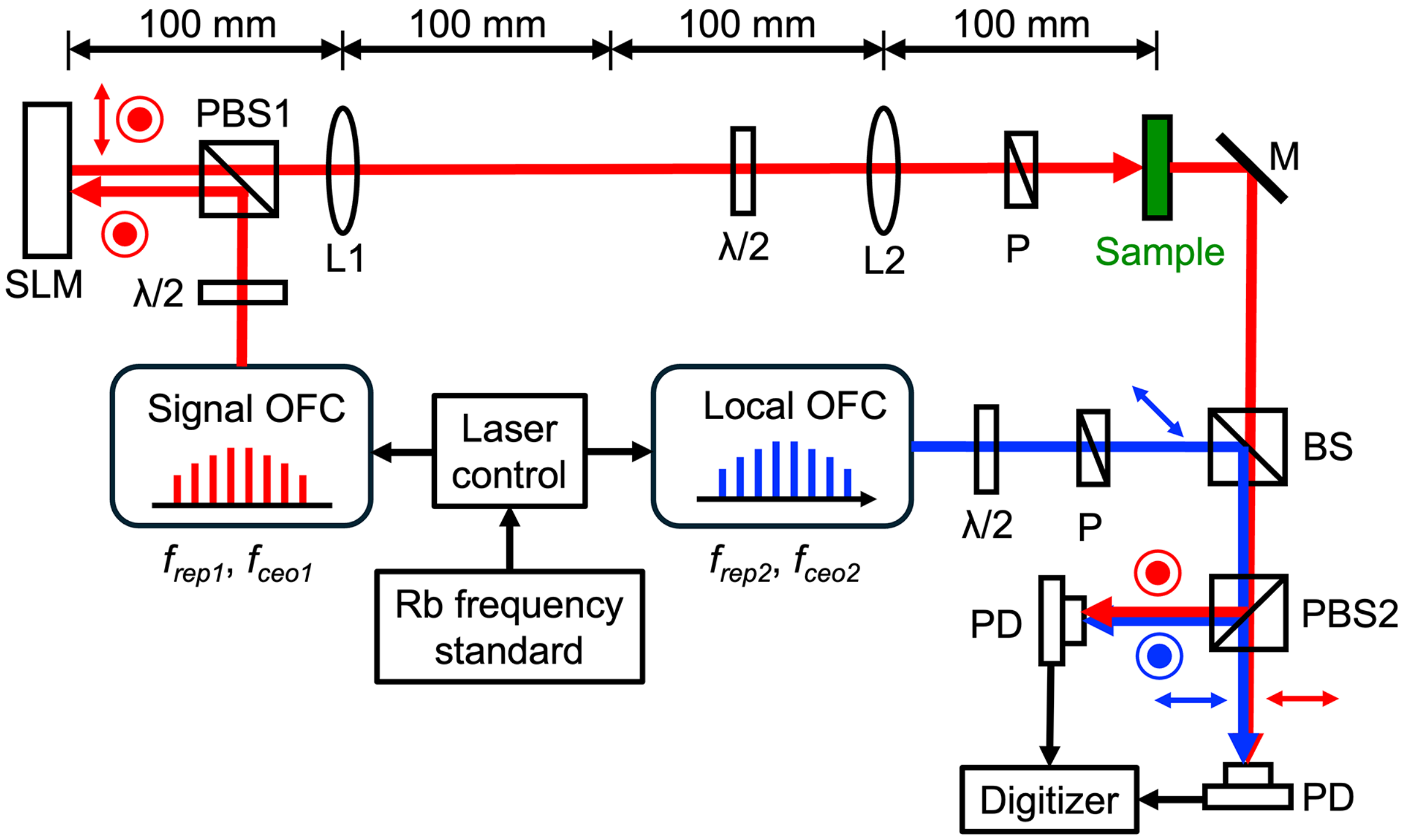

Fig. 2. Experimental setup of the proposed dual-comb spectro-polarimetric single-pixel imaging (DCSP-SPI) system. The system consists of three modules: optical frequency comb (OFC) sources, an SLM unit for structured illumination, and a DCSP detection unit. The signal OFC is spatially modulated by a liquid-crystal-on-silicon spatial light modulator (LCOS-SLM) to generate Hadamard-based structured illumination patterns, which are relay-imaged onto the sample. The transmitted light is interfered with the local OFC, separated into two orthogonal polarization components, and simultaneously detected by dual photodetectors. Fourier transformation of the dual-comb interferograms yields polarization-resolved optical amplitude and phase spectra, from which the amplitude ratio ($\psi$) and phase difference ($\Delta$) are calculated. Correlation of the spectro-polarimetric data with the structured illumination patterns reconstructs hyperspectral images of optical amplitude, optical phase, $\psi$, and $\Delta$.

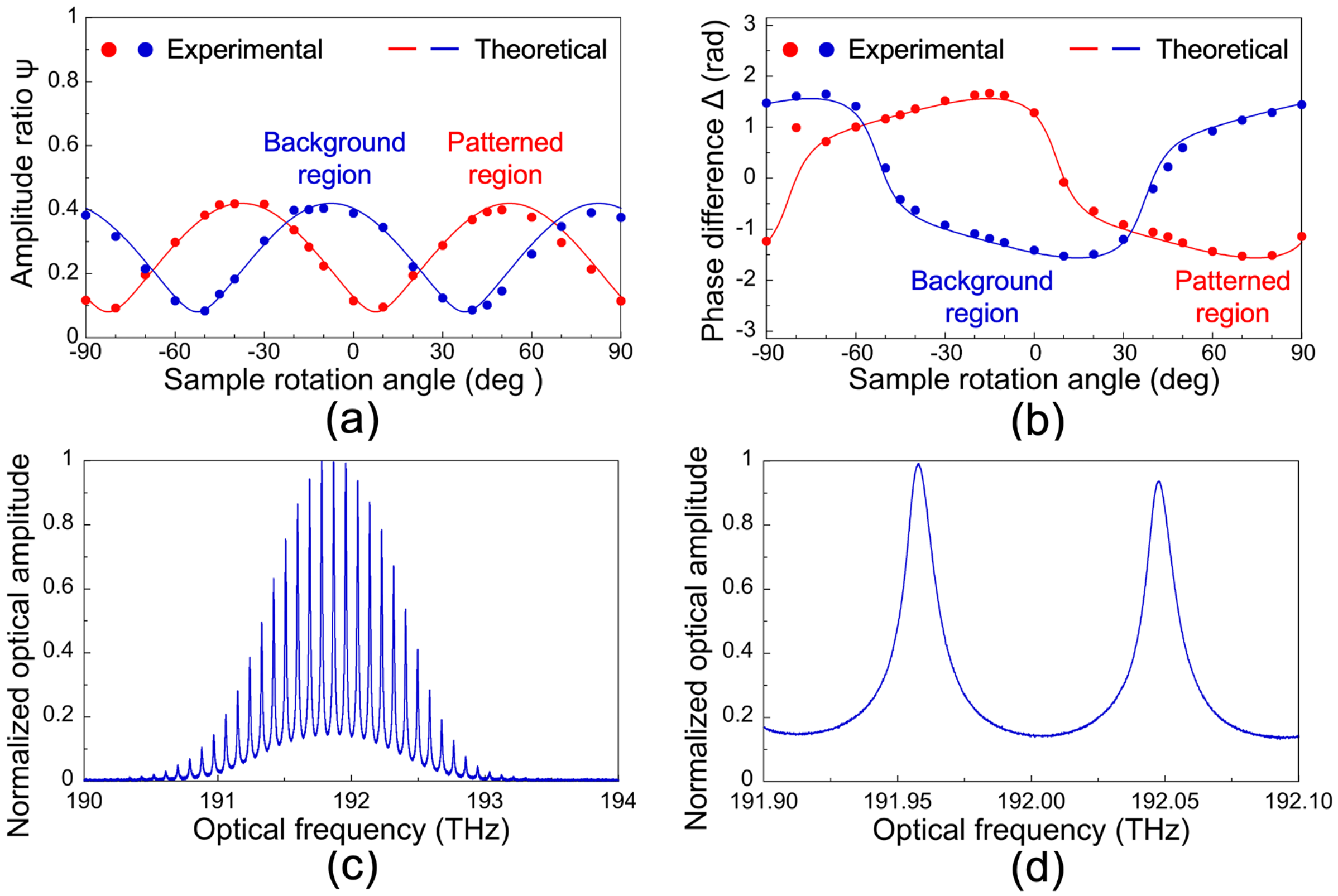

Fig. 3. Polarization and spectral characteristics of the sample used for DCSP-SPI measurements. (a) Measured and theoretical amplitude ratio ($\psi$) and (b) phase difference ($\Delta$) as functions of the sample rotation angle for the background and patterned regions of the birefringent resolution target. (c) Transmission spectrum of the Fabry–Pérot etalon filter used to provide wavelength-dependent spectral modulation. (d) Magnified view of the transmission spectrum around 192 THz, showing two representative transmission peaks used for spectro-polarimetric imaging.

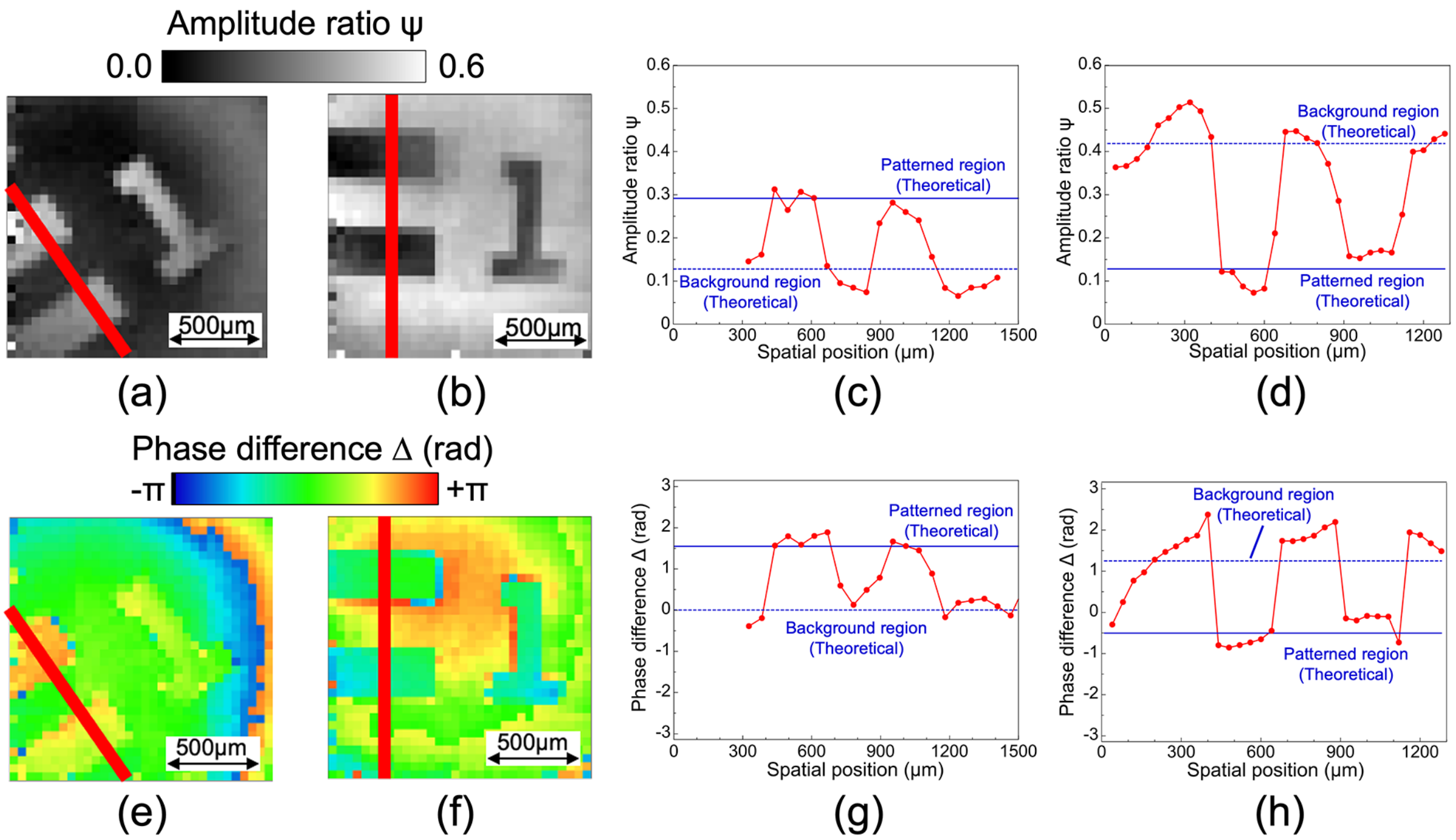

Fig. 4. Single-frequency spectro-polarimetric single-pixel imaging of a birefringent resolution target at an optical frequency of 192.0125 THz. (a,b) Reconstructed amplitude-ratio ($\psi$) images for sample rotation angles of 10° and 45°, respectively. (c,d) Corresponding line profiles of $\psi$ extracted along the red lines in (a) and (b), respectively. Red plots represent the experimental data, while the blue solid and dashed lines indicate the theoretical values. (e,f) Reconstructed phase-difference ($\Delta$) images for sample rotation angles of 10° and 45°, respectively. (g,h) Corresponding line profiles of $\Delta$ extracted along the red lines in (e) and (f), respectively. Red plots represent the experimental data, and the blue solid and dashed lines denote the theoretical values.

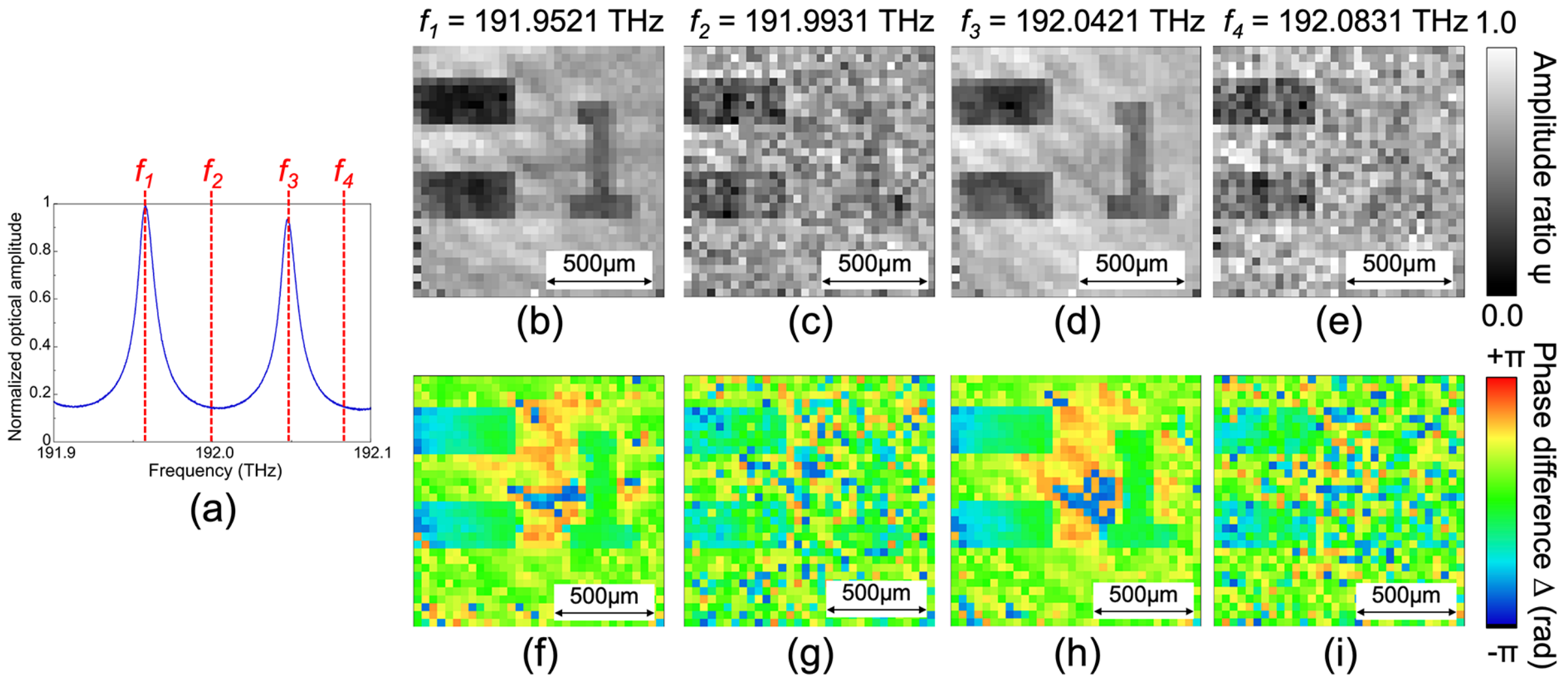

Fig. 5. Hyperspectral spectro-polarimetric single-pixel imaging of a birefringent resolution target combined with a Fabry–Pérot etalon. (a) Normalized optical amplitude spectrum of the Fabry–Pérot etalon, indicating the four representative optical frequencies ($f_1$–$f_4$) selected for hyperspectral imaging. (b–e) Reconstructed amplitude-ratio ($\psi$) images at optical frequencies of (b) $f_1$ = 191.9521 THz, (c) $f_2$ = 191.9931 THz, (d) $f_3$ = 192.0421 THz, and (e) $f_4$ = 192.0831 THz. The frequencies in (b) and (d) correspond to transmission peaks of the Fabry–Pérot etalon, whereas those in (c) and (e) correspond to transmission minima. (f–i) Corresponding phase-difference ($\Delta$) images at the same optical frequencies. The image signal-to-noise ratio varies with the etalon transmission, while the birefringent features are consistently preserved in both $\psi$ and $\Delta$ images.

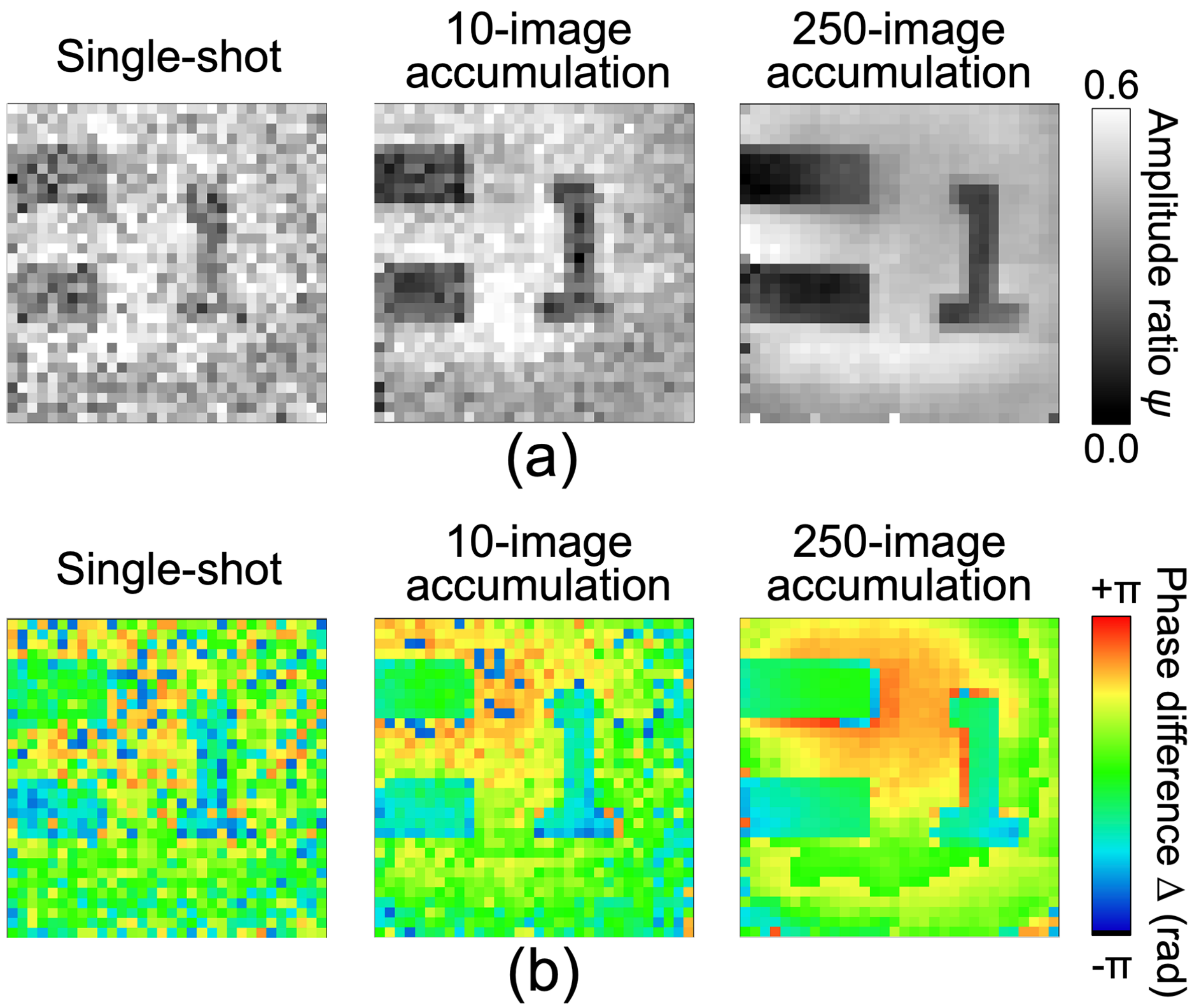

Fig. 6. Effect of coherent accumulation on the image quality of DCSP-SPI. (a) Reconstructed amplitude-ratio ($\psi$) images obtained using single-shot acquisition, 10-image coherent accumulation, and 250-image coherent accumulation. (b) Corresponding phase-difference ($\Delta$) images reconstructed under the same acquisition conditions. Increasing the number of coherently accumulated interferograms progressively improves the image signal-to-noise ratio while preserving the underlying spatial contrast.

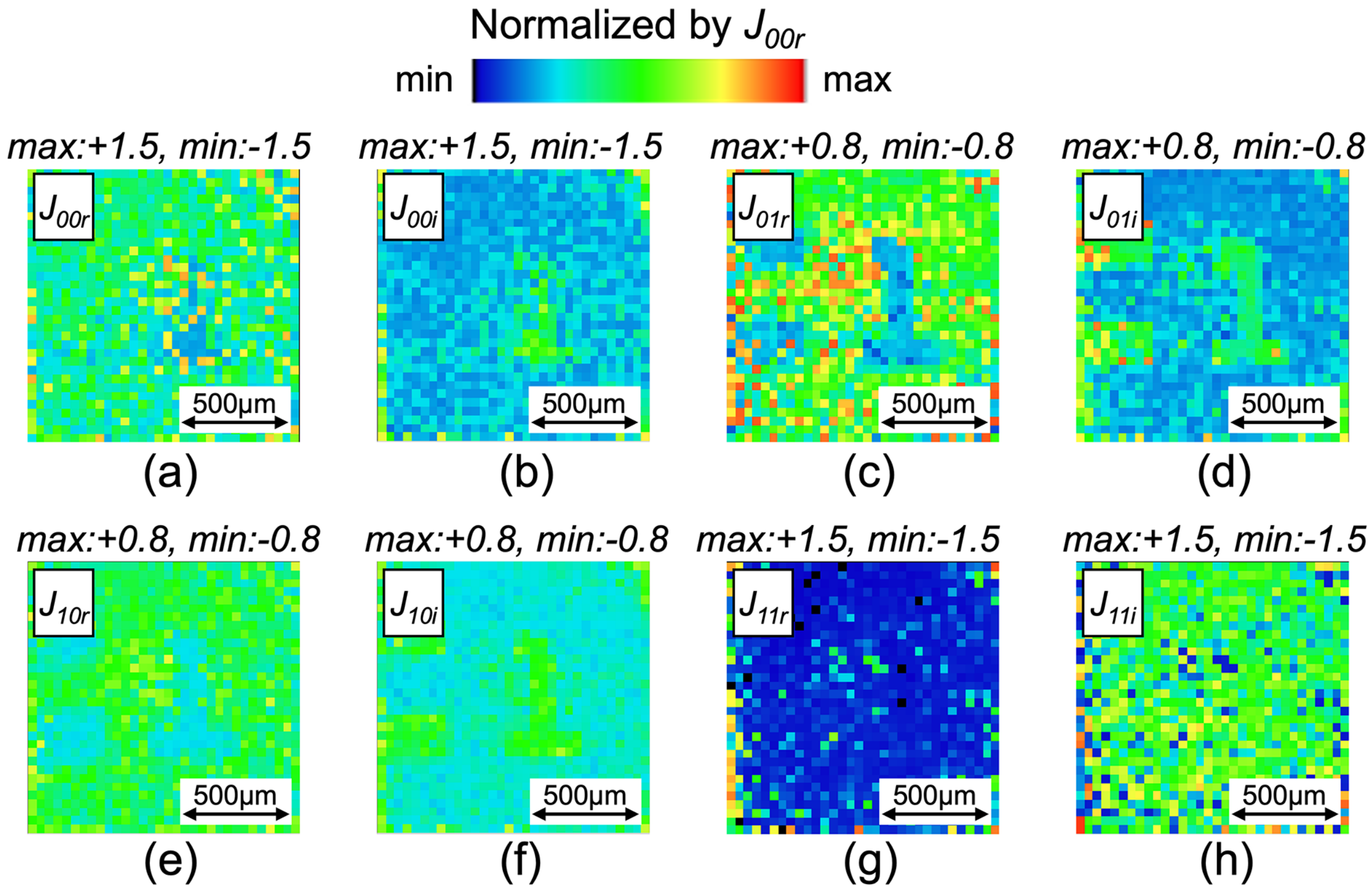

Fig. 7. Hyperspectral Jones matrix single-pixel imaging of a birefringent resolution target combined with a Fabry–Pérot etalon at an optical frequency of 192.0421 THz, corresponding to a transmission peak of the etalon. (a–h) Reconstructed real and imaginary parts of the Jones matrix elements: (a) Re($J_{00}$), (b) Im($J_{00}$), (c) Re($J_{01}$), (d) Im($J_{01}$), (e) Re($J_{10}$), (f) Im($J_{10}$), (g) Re($J_{11}$), and (h) Im($J_{11}$). All images are normalized to the maximum value of Re($J_{00}$). The reconstructed Jones matrix exhibits distinct spatial contrasts corresponding to the birefringent sample, while the close agreement between the off-diagonal elements $J_{01}$ and $J_{10}$ confirms the validity of the hyperspectral Jones matrix reconstruction.

Table 1. Estimated measurement time of the proposed DCSP-SPI system using LCOS-SLM and Digital Micromirror Device (DMD) implementations under different numbers of coherent accumulations. The total measurement time was estimated from the interferogram acquisition time, pattern switching time, and the number of Hadamard patterns.

<table>
<tr><th>Spatial light modulator</th><th>Number of coherent accumulations</th><th>Interferogram acquisition time</th><th>Accumulated interferogram acquisition time</th><th>Pattern switching time</th><th>Number of Hadamard patterns</th><th>Total measurement time</th></tr>
<tr><td rowspan="3">LCOS-SLM</td><td>250</td><td rowspan="3">9.5 ms/interferogram</td><td>2.4 s/pattern</td><td rowspan="3">2.2 s/pattern</td><td rowspan="3">1024</td><td>4710 s</td></tr>
<tr><td>10</td><td>95 ms/pattern</td><td>2350 s</td></tr>
<tr><td>1</td><td>9.5 ms/pattern</td><td>2263 s</td></tr>
<tr><td rowspan="3">Digital Micromirror Device</td><td>250</td><td rowspan="3">9.5 ms/interferogram</td><td>2.4 s/pattern</td><td rowspan="3">240 μs/pattern</td><td rowspan="3">1024</td><td>2458 s</td></tr>
<tr><td>10</td><td>95 ms/pattern</td><td>98 s</td></tr>
<tr><td>1</td><td>9.5 ms/pattern</td><td>10 s</td></tr>
</table>